\documentclass[11pt]{article}

\usepackage{amsmath}
\usepackage{amssymb}
\usepackage{amsthm}

\usepackage{graphicx}

\graphicspath{{}} % define the path to the folder with the figures

\usepackage{cite}

\usepackage[labelfont=bf,labelsep=period,justification=raggedright]{caption}
\usepackage{epsfig}
\everymath{\displaystyle}

\makeatletter
\renewcommand{\@biblabel}[1]{\quad#1.}
\makeatother

\date{}

\newcommand*\Laplace{\mathop{{}\bigtriangleup}\nolimits}

\begin{document}

\title{
Weber's law based perception and the stability of animal groups
}
\author{
Andrea Perna$^{1,2}$,
Giulio Facchini$^{1}$,
Jean Louis Deneubourg$^{2}$
\\
\textit{$^{1}$ Life Sciences Department, University of Roehampton, London, UK}\\
\textit{$^{2}$ Unit of Social Ecology, Université Libre de Bruxelles, Bruxelles, Belgium}
}

\maketitle
\begin{abstract}
Group living animals form aggregations and flocks that remain cohesive in spite of internal movements of individuals. This is possible because individual group members repeatedly adjust their position and motion in response to the position and motion of other group members. Here we develop a theoretical approach to address the question, what general features -if any- underlie the interaction rules that mediate group stability in animals of all species? We do so by considering how the spatial organisation of a group would change in the complete absence of interactions. Without interactions, a group would disperse in a way that can be easily characterised in terms of Fick's diffusion equations. We can hence address the inverse theoretical problem of finding the individual-level interaction responses that are required to counterbalance diffusion and to preserve group stability. We show that an individual-level response to neighbour densities in the form of Weber's law (a `universal' law describing the functioning of the sensory systems of animals of all species) results in an `anti-diffusion' term at the group level. On short time scales, this anti-diffusion restores the initial group configuration in a way which is reminiscent of methods for image deblurring in image processing. We also show that any non-homogeneous, spatial density distribution can be preserved over time if individual movement patterns have the form of a Weber's law response. Weber's law describes the fundamental functioning of perceptual systems. Our study indicates that it is also a necessary -but not sufficient- feature of collective interactions in stable animal groups.

\begin{center}
\end{center}
\end{abstract}

\section{Introduction}
Gregarious and social animals are capable of forming aggregations and flocks that maintain a relatively stable configuration in spite of individual animals joining or leaving the group and continuously changing their position relative to each other. Examples of such aggregations span the whole animal kingdom from molluscs (e.g. \cite{vandeKoppel2008}), to insects (e.g. \cite{Puckett2014}) and crustacea (e.g. \cite{Devigne2011woodlice}) to vertebrates (reviewed in \cite{Krause_and_Ruxton2002}). All these groups can remain cohesive and maintain a stable, coherent organisation over time because each member of the group continuously tracks the position or density of its surrounding neighbours and implements appropriate `interaction responses' that effectively preserve group stability. 

A large number of recent empirical studies has aimed at identifying these individual-level interaction responses empirically from tracking data of animals of various species (e.g. \cite{Ballerini2008,Herbert-Read2011,Gautrais2012,Strandburg-Peshkin,Attanasi2014,Herbert-Read_escape_waves2015}). Together, these studies have supported the idea that group formation requires at least some form of inter-individual attraction. However, different studies have pointed to different attraction rules in different species. For instance attraction could be directed to one single individual at a time or simultaneously to multiple neighbours; in turn, these relevant neighbours could be selected based on metric or topological distance, etc.

The problem is compounded by the fact that the interaction responses of animals of different species ultimately depend on the sensory modalities mediating the interaction (vocal calls, vision, pheromones), and on the underlying neural circuits involved, which are inevitably different from one species to another. In fact, there is a large variability across the sensory systems of different animals, with only very few 'perceptual laws' that are shared both across taxa and across sensory modalities. One such `universal' perceptual law is the Weber-Fechner law \cite{Weber1834}, which states that the ability of a sensory system to discriminate between two physical quantities decreases in inverse proportion to the magnitude of the quantities being compared. To make an example, we can easily tell the difference between a cluster of five objects and a cluster of eight, but we cannot as easily identify the difference between -say- a cluster of 55 and one of 58 \cite{anobile2016number}. Weber's law has been established for humans as well as for a wide range of animals ranging from ants \cite{von2014pheromone}, to fish\cite{Agrillo2008}, to corvids \cite{Ditz2016}, and to primates\cite{jordan2006weber} and occurs in different sensory modalities. Because Weber's law controls the ability of animals to `count' the number of neighbours in different directions, it is likely to play a role also in collective aggregation phenomena.

Here, we aim at investigating the general properties of the interaction responses mediating stability in animal groups. Whereas many of the previous studies have tackled this problem starting from the analysis of tracking data, here we develop a complementary theoretical approach in which we start by considering the theoretical problem of an animal group whose member individuals \emph{do not} interact. In the absence of interactions, the dynamics of the group is dominated by diffusion processes which can be accurately characterised in terms of Fick's diffusion equations \cite{Fick1855} under simple assumptions of random motion and absence of interactions. We can hence address the inverse theoretical problem of finding which individual-level interaction rules are required to precisely counter-balance the effects of diffusion and preserve group stability.

% Results and Discussion can be combined.
\section{Evolution of densities in the absence of interactions}
In the absence of interactions, the movement in random directions of individual animals results in a change of densities over time that is well described by the diffusion equation \cite{Fick1855}. Here we recall it briefly for the one dimensional case (the problem is analogous in a higher dimensional space).

\begin{figure*}[bt]
\centering
\includegraphics[width=0.7\textwidth]{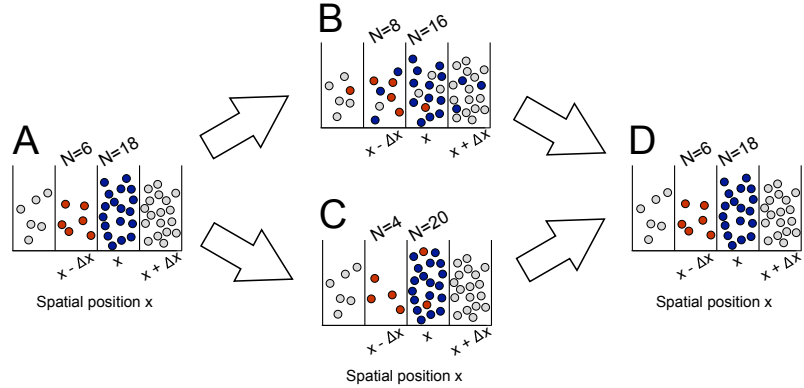}
\caption{{\bf Diffusion and anti-diffusion in a discrete space.} (\textbf{A} and \textbf{B}) Diffusion: each particle taking part in the initial distribution illustrated in (A) has a constant probability $D$ of moving to the adjacent cell on the left, and an equal probability of moving to the cell on the right. If -as it is the case in this example- the probability is set to $D = 1/6$, three particles will move on average from cell $x$ to cell $x - \Delta x$ and on average one particle will move in the opposite direction from $x - \Delta x$ to $x$. While individual particles do not show any directional preference, and they do not interact in any way, the overall process produces (in this example) a net flow of two particles from the region of higher concentration $x$ to the region of lower concentration $x - \Delta x$. (\textbf{A} and \textbf{C}) Anti-diffusion based on Weber's Law: here each particle has a probability of moving against the concentration gradient that is proportional to the Weber's fraction of concentrations. If the proportionality constant is set to $\gamma = 1/6$, two particles on average will move from $x - \Delta x$ to $x$. (\textbf{D}) If both diffusion and anti-diffusion take place simultaneously the net effect is that of stabilising group configuration.}
\label{fig:diffusion_in_discrete_space}
\end{figure*}

For convenience, we imagine the space to be subdivided in cells of equal width $\Delta x$ and we indicate with $C(x,t)$ the density of individuals (e.g. number of individuals over volume of the cell) at the position $x$ and time $t$ (fig. \ref{fig:diffusion_in_discrete_space}). We focus on density, not on the absolute number of individuals, so that our quantities do not scale with the size of the cells.
Within a certain time interval $\Delta t$ a fraction $2D$ of individuals present in each cell move to one randomly chosen adjacent cell. This determines the following equation for the evolution of $C$:
\begin{equation}
\begin{split}
C(x, t+\Delta t)  = \\
C(x, t) + D \left[- 2 C(x,t) + C(x+\Delta x,t) + C(x -\Delta x,t) \right]
\label{eq:diffusion_discrete}
\end{split}
\end{equation}
where the density at some position $x$ depends on the previous density at the same position, minus the fraction of individuals that moved to the neighbour cells, plus the individuals that moved into the cell from the neighbouring ones. In the limit for $\Delta t$ and $\Delta x$ small, equation \ref{eq:diffusion_discrete} becomes
\begin{equation}
\frac{\partial C}{\partial t} = D' \frac{\partial^2 C}{\partial x^2}
\label{eq:diffusion_continuous}
\end{equation}
where $D' = D\frac{\left(\Delta x\right)^2}{\Delta t}$ (see Electronic Supplementary Material for a derivation).

The solution of the diffusion equation is
\begin{equation}
C(x, t) = C(x,0) \ast G(0, \sqrt{t})
\label{eq:diffusion_solved}
\end{equation}
whereby the initial distribution $C(x, 0)$ is convolved with a Gaussian $G(0, \sqrt{t})$ that becomes larger and larger in time.
% The explicit form of $G$ is $\frac{1}{\sigma_d(t) \sqrt{2 \pi}} e^{- \frac{x^2}{\textcolor{red}{2} \sigma_d(t)^2}}$, with $\sigma_d(t) = \sqrt{2 D' t}$.
In practice, it is as if the initial distribution was replaced by an increasingly `blurred' version of itself.

% \section*{Restoring density distributions after diffusion}
\section{Individual-level responses based on Weber's law produce collective-level anti-diffusion}
In order to contrast diffusion, the individuals at each given location $x$ must have an increased tendency to move to an adjacent location $x + \Delta x$ if the concentration of neighbours at the target location is higher than the concentration at $x$.  Weber's law states that the probability to discriminate the higher concentration $C(x+\Delta x)$ from the local concentration $C(x)$ is a monotonically increasing function $F$ of the normalised concentration difference:
\begin{equation}
F\left(\frac{ \left| C(x+ \Delta x) - C(x) \right| }{C(x)}\right)
\label{eq:monotonicFunction}
\end{equation}
In the context of the discrimination between sensory stimuli, the argument of $F$ is often indicated as the Weber fraction of stimulus intensities. \\
Let's consider here how this type of response would affect the evolution of local densities in the same one-dimensional discretised example of figure \ref{fig:diffusion_in_discrete_space} by focusing on the following model to describe the probability of moving between $x$ and $x + \Delta x$. 
\begin{equation}
p(x\rightarrow x+\Delta x)
\begin{cases}
\gamma \frac{ \left| C(x+ \Delta x) - C(x) \right| }{C(x)}, & \textnormal{if}\: C(x + \Delta x) > C(x)\\
0 & \textnormal{otherwise}
\end{cases}
\label{eq:individualWeber}
\end{equation}
In our specific model, the probability of response is simply linear with the normalised concentration difference ($F$ is simply replaced by the proportionality constant $\gamma$) and we do not include explicitly any noise term. We choose this particular formulation because of its nice properties in relation to the diffusion equation, which will become apparent below. In a perfectly biologically realistic scenario,  probabilities of response cannot increase linearly (for instance they cannot become larger than one) and psychophysically measured response thresholds often have a logistic shape, so our approximation is only accurate when the argument of $F$ is small. In supplementary section S-4 we also consider briefly the case of non-linear $F$.

While we do not include an explicit noise term in equation \ref{eq:individualWeber}, noise is accounted by the fact that individuals undergo diffusion, as well as they move up concentration gradients (for example, the probability of actively moving in response to the concentration gradient becomes zero when $C(x + \Delta-x) = C(x)$, but individuals can still move in both directions through `diffusion').

The effect on local concentrations of individuals produced by equation \ref{eq:individualWeber} can easily be estimated by considering that there
are $C(x)$ individuals that apply the gradient-response rule. This will result in a net flow from $x$ to $x + \Delta x$ equivalent to $\gamma \left(C(x+\Delta x) - C(x) \right)$ if $x + \Delta x$ is the cell with higher concentration. If instead the cell with higher concentration is the one at $x$, particles will flow in the opposite direction, but their flow will turn out to be also proportional to $\gamma \left(C(x+\Delta x) - C(x) \right)$. As a result the net change of density at $x$ will be:\\
\begin{equation}
\begin{split}
C(x,t+\Delta t) = \\
C(x, t) - \gamma \left[ - 2 C(x,t) + C(x + \Delta x, t) + C(x - \Delta x, t)\right]
\label{eq:antidiffusion_discrete}
\end{split}
\end{equation}
whose continuous version is:
\begin{equation} 
\frac{\partial C}{\partial t} = - \gamma' \frac{\partial^2 C}{\partial x^2}
\label{eq:antidiffusion_continuous}
\end{equation}
%
% \begin{equation}
% p(x_i \rightarrow x_j) = \gamma \frac{C_j - C_i}{C_i}
% \end{equation}
where $\gamma' = \gamma \frac{\left(\Delta x\right)^2}{\Delta t}$.

The Electronic Supplementary Material provides another intuitive justification for the choice of focusing on Weber's law in the case of normally distributed densities).

Equations \ref{eq:antidiffusion_discrete} and \ref{eq:antidiffusion_continuous} are identical to the diffusion equations \ref{eq:diffusion_discrete} and \ref{eq:diffusion_continuous}, except for the sign of the diffusion coefficients $D$ and $\gamma$. Importantly, however, their biological justification is completely different: the diffusion equation results from the total absence of interactions, while a response based on Weber's law implies an active decision process (see also \cite{keller1971traveling} where a similar `anti-diffusion' term was obtained, although in a slightly different context). %, but whose effect (assuming individuals decide to move down the gradient) would be indistinguishable from that of diffusion in the absence of any decision!

Care should be taken however that the derivation of the anti-diffusion equation \ref{eq:antidiffusion_discrete} from individual-level Weber's law responses (eq. \ref{eq:individualWeber}) involves some simplifications. First, equation \ref{eq:antidiffusion_discrete} is a mean field approximation of equation \ref{eq:individualWeber}. Second, if the number of particles or individuals in a cell is close to zero, the probability for these particles to move to an adjacent higher density cell will increase (the denominator of equation \ref{eq:individualWeber} is small), but the flow can never exceed the number of available particles. There is no control for this in equation \ref{eq:antidiffusion_discrete}, which means that simply applying this equation can potentially lead to some cells taking negative values.

\begin{figure}[tbp]
\centering
\includegraphics[width=1\columnwidth]{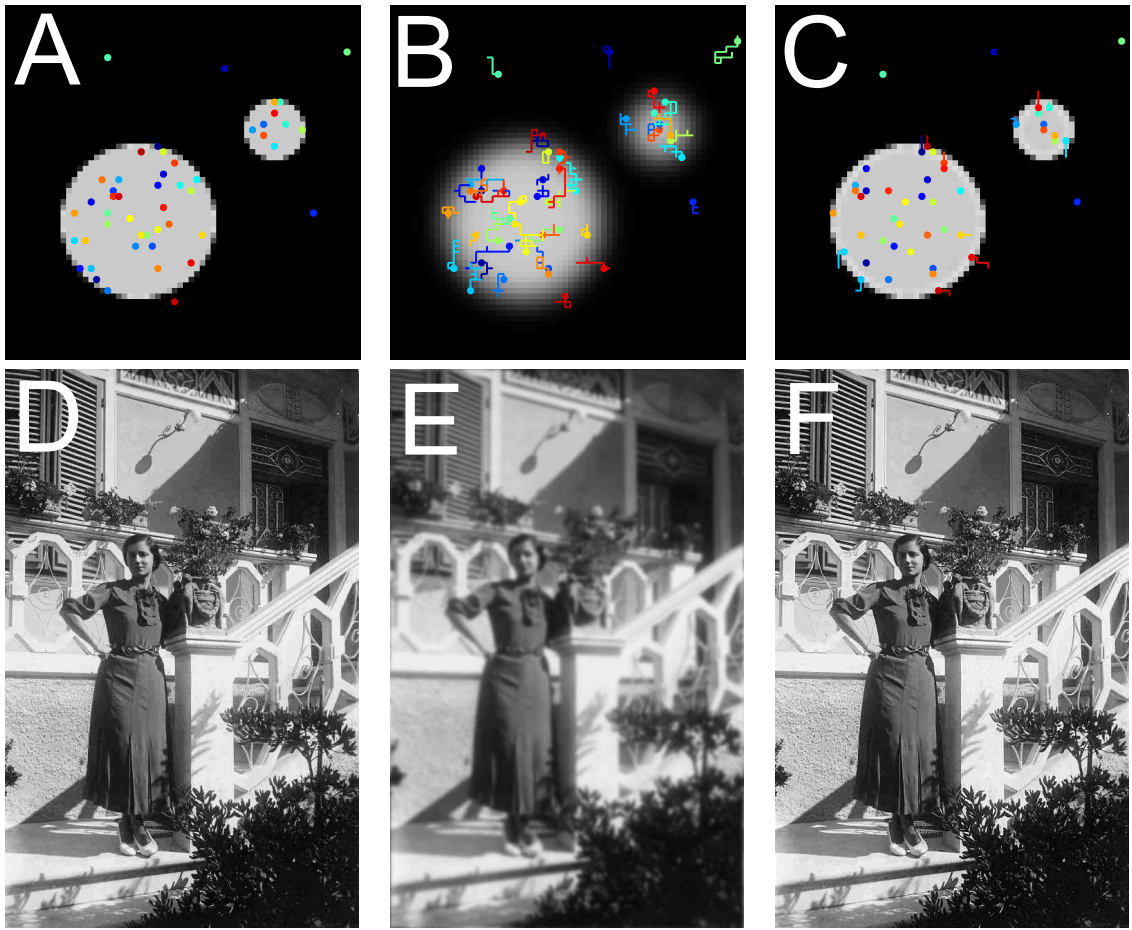}
\caption{
{\bf Weber's law based anti-diffusion restores spatial distributions after diffusion.} \textbf{A} Hypothetical two-dimensional density distribution representing two 'animal groups' with a circular profile and homogeneous internal density. Brighter regions indicate higher density. The coloured dots indicate the position of randomly selected individuals. After a given number of iterations of random diffusion-like movements (300 in this example), individuals have moved with random trajectories producing a smoothed density distribution (\textbf{B}). In \textbf{C} a density distribution similar to the original (A) distribution is restored by each individual climbing the density gradient with a Weber's law response rule (parameter $\gamma = D$ and same number of iterations: 300). Figures \textbf{D}, \textbf{E} and \textbf{F} illustrate an identical process to that in A, B and C, but here the starting density distribution (\textbf{D}) is a grey-scale image. Random diffusion of the individuals that form the distribution in D has the effect of blurring the image (producing the image in \textbf{E}) and anti-diffusion restores a similar image to the original (in \textbf{F}). In both examples the luminance map is updated based on a mean-field approach (that is, densities are updated in proportion to the calculated probability instead of by actually drawing movement decisions from the probability distribution).
}
\label{fig:diffusion_and_antidiffusion}
\end{figure}

\textbf{Restoring arbitrary distributions} \\
Because a response to concentration gradients based on Weber's law results in a change in concentrations that is analogous to an anti-diffusion, if such a response is implemented by all individuals in a group, it will have the net effect of partially restoring the distribution that was originally altered by diffusion. To visualise this, consider the example in the top row of figure \ref{fig:diffusion_and_antidiffusion}. The example depicts the hypothetical case of animals aggregated at two high density spots, with densities close to zero elsewhere (fig. \ref{fig:diffusion_and_antidiffusion}-A). Coloured dots mark the positions of randomly selected particles. In the absence of interactions, diffusion operates on densities, which for the purpose of figure \ref{fig:diffusion_and_antidiffusion} we implemented by iterating 300 times a 2D version of equation \ref{eq:diffusion_discrete} with parameter $D = 0.01$. The superposed particles also move to a random cell adjacent to their current position with probability $D$. Figure \ref{fig:diffusion_and_antidiffusion}-B shows the resulting density distribution and the trajectories of the selected particles. 
Applying 300 iterations of anti-diffusion response (a 2D version of equation \ref{eq:antidiffusion_discrete}) results in a new spatial distribution (fig. \ref{fig:diffusion_and_antidiffusion}-C) where densities are very similar to those in the original distribution.

{\bf An analogy with the problem of deblurring in image processing}

Our problem of animals recovering a group configuration that was previously altered by diffusion is closely related to the problem of deblurring in image processing. In fact, a grey-scale image can be seen as a density distribution (for instance by assimilating bright colours to high density regions and dark colours to low density regions). Random movements starting from that distribution correspond to blurring the image, and the task of image deblurring algorithms is that of recovering an image similar to the original one. It is well known in the human and computer vision literature that blurred images can be partially restored by subtracting their convolution with a Laplacian operator, in a similar way to what equation \ref{eq:antidiffusion_discrete} does. In fact, the diffusion equation \ref{eq:diffusion_continuous} indicates that the local change in luminance $\partial C$  when we blur the image by a small amount $\partial t$ is proportional to the Laplacian of the original image: 
\begin{equation} 
\frac{\partial C}{\partial t} = D' \left[ \frac{\partial^2 C}{\partial x^2} + \frac{\partial^2 C}{\partial y^2} \right]
\label{eq:laplacian2D}
\end{equation}
A simple approach for restoring the quality of the initial image is that of playing the diffusion process for negative time scales \cite{terHaarRomeny1993} in a way which is analogous to what is done by equation \ref{eq:antidiffusion_discrete}.

Figure \ref{fig:diffusion_and_antidiffusion} D-E illustrates the diffusion and Weber's law based anti-diffusion applied to a grey-scale image. The original image (fig.  \ref{fig:diffusion_and_antidiffusion}-D) is initially blurred when each of the intensity levels of a pixel randomly diffuses to the neighbouring pixels (fig.  \ref{fig:diffusion_and_antidiffusion}-E). Implementing a Weber's law based gradient-climbing response such as the one described in equation \ref{eq:individualWeber} restores much of the details that were present in the original image. The only difference between this gradient-climbing approach and classical image deblurring by subtraction of a laplacian convolution is that our approach is based on equation \ref{eq:individualWeber}, while the convolution with a laplacian implements directly equation \ref{eq:antidiffusion_discrete}, which can be derived from it. Unlike gradient climbing, subtracting the laplacian convolution can produce images with negative values; in the context of image processing this is usually resolved by rescaling the image histogram to the allowed range of pixel intensities (e.g. 0-255 for 8-bit images).

% One of the properties of the convolution is that the derivative of the convolution of two functions is equal to the convolution of the first function with the derivative of the second one: $(f \ast g)' = f \ast g'$. If a blurred image is the convolution of a sharp image with a gaussian, then we are sure that the blurred image is differentiable infinitely many times, because the Gaussian is differentiable even if the starting image wasn't.

\section{Animal movements within stable density landscapes}
The previous examples focused on a somewhat artificial situation whereby particles initially diffuse without interacting and then converge again by implementing a Weber's law based gradient climbing response. 

More realistic is the situation in which a group or a population of animals occupy the environment according to a non-homogeneous density distribution which remains stable over time in spite of individual animals moving in different directions. 

% \subsection*{Internal movements within an arbitrary probability landscape}
% MATLAB: simulate_stable_landscape.m
We model this situation by assuming that animals occupy the environment according to a non-homogeneous density landscape, that for simplicity we map to a discrete lattice. Animals are simulated in the model as particles that can move from one cell of the lattice to the adjacent cells, but we impose that the density landscape remains unchanged (see Electronic Supplementary Material for details).

\begin{figure}[tb]
\centering
\includegraphics[width=1\columnwidth]{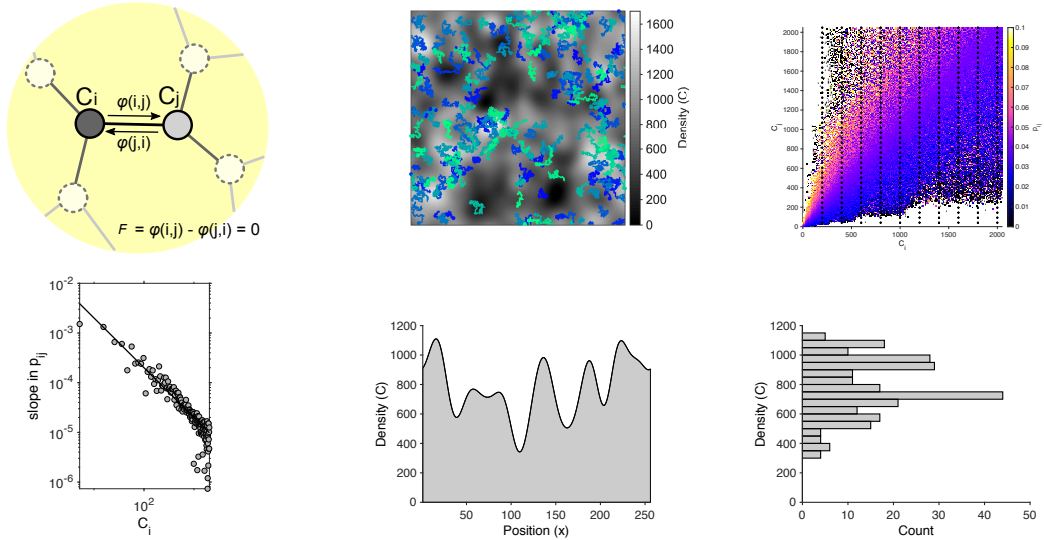}
%\begin{minipage}[t]{1\textwidth}
%\begin{tabular}{cc}
%{\includegraphics[width=0.3\textwidth]{figure_network}} &
%{\includegraphics[width=0.4\textwidth]{traj00500}} \\
%{\includegraphics[width=0.5\textwidth]{probVsNiNjwithLines}} &
%{\includegraphics[width=0.5\textwidth]{slopePvsNi}} \\
%\end{tabular}
%\end{minipage}
\caption{
{\bf Flow of individuals across a network of `sites'}. \textbf{(a)} The simplest way in which a non-flat density landscape can remain stable over time is if the flows of individuals moving across each edge in one direction is balanced by the flow moving in the opposite direction. We built a series of density landscapes such as in \textbf{(b)} whereby each pixel represents a network node and it is directly connected by an edge to its four adjacent pixels. Particles move randomly across the landscape with the constraints that the density distribution is preserved at all times. \textbf{(c)} The probability for a particle to move from node $i$ to node $j$ increases linearly with the difference of density between the target $C_j$ and the source node $C_i$. The slope of this relation, obtained from a linear fit along the dotted lines depicted in (c) is inversely proportional to the local density $C_i$ (panel \textbf{(d)}). A spatial distribution of densities \textbf{(e)} can also be summarised in terms of its histogram \textbf{(f)}, and a possible alternative definition of group stability is one in which the histogram is conserved over time, while the distribution itself can change.  
}
\label{fig:network}
\end{figure}

Suppose that the number of individuals at two adjacent cells $i$ and $j$ is respectively $C_i$ and $C_j$ (fig. \ref{fig:network}). If the density landscape remains stable, $C_i$ and $C_j$ must remain constant over time (and this for all cells, not just for $i$ and $j$). The easiest way in which this can be achieved (excluding the contribution of flows `running in circle' over longer loops) is if at any given time the flow $\phi (i,j)$ from $i$ to $j$ is equal to the flow $\phi(j,i)$ in the opposite direction. % I could not show that this is the only possible way to maintain a balanced density (with very specific exceptionsa of flat distributions etc.)

Assuming that $j$ is the cell with higher density $C_j > C_i$, we can say that the individuals that move from $j$ to $i$ do so because of diffusion alone, while the individuals that move from $i$ to$j$ can do it because of diffusion or because of some form of gradient climbing described by an \textit{a priori} unknown function of local densities $f(C_i, C_j)$.

% \ref{sec:Gaussian_stable_landscape}
% \textcolor{red}{Here we are saying that if the distribution remains constant the flow across each edge in both directions must be balanced. In a portion of network without loops, this is clearly true. When there are loops things get more difficult, though in general I think it remains the case that $\phi (i,j) = \phi (j,i)$: if the flows clockwise and counterclockwise around the loop were different, as soon as there are different numbers on different nodes the difference would also propagate clockwise or counterclockwise. The balance cannot come from }
% \textcolor{red}{We know that the sum of individuals entering a node through all adjacent edges must equal the sum of individual leaving the same node (that is, first Kirchhoff law holds) and that we have $X = PX$, where $X$ is the number of individuals in each node and $P$ is the transition matrix (which in general is a function of $X$, but here we consider the case of $X$ constant).}

We then have 
\begin{equation}
\phi (i,j) = C_i D + C_i f(C_i, C_j) = C_j D = \phi(j,i)
\end{equation}

which gives 

\begin{equation}
f(C_i, C_j) = \frac{D (C_j - C_i)}{C_i}
\label{eq:Weber_balancing_flows}
\end{equation}

Weber's law is hence a solution.

We tested in simulation if particles moving within an overall stable density landscape appear as if they were following Weber's law response to densities. First, we created a number of 2-dimensional random density landscapes (obtained as low-pass filtered noise). Then, we generated random particle trajectories with the double constraint that (i) the total amount of time spent by a trajectory at a particular location is proportional to the density at that location, and that (ii) at each time step the net flow of particles between two locations is equal to the flow in the opposite direction (see Electronic Supplementary Material for details; an example of one such density landscape and of some of the corresponding trajectories is illustrated in figure \ref{fig:network}(b)).
% This is the imposed density distribution. In order to have reasonable transition probabilities for this particular density distribution we look for a matrix of transition probabilities for which the stationary distribution is an eigenvector. 
The probability for particles to move between two adjacent locations calculated over all the trajectories and binned as a function of the local densities at the source ($C_i$) and at the target location ($C_j$) is illustrated in figure \ref{fig:network}(c). While the flow of particles is identical in both directions, individual particles necessarily have a higher probability of moving from low to high density regions. Intuitively, this is easily understood as a consequence of the fact that fewer particles are available in lower density cells, and as a consequence they are required to have a higher probability of moving to counterbalance the flow from higher density cells.
More precisely, the simulated particles responded to local densities with an apparent Weber's law response, whereby the slope of the probability of moving from $i$ to $j$ is inversely proportional to $C_i$ (fig. \ref{fig:network}(d)).
These simulations show that as long as the overall spatial density distribution is maintained, individual group members will appear to respond to their neighbours with a Weber's law type response.

In our case scenario of a stable density landscape, a gradient response compatible with Weber's law results from the requirement of balancing flows for each pair of adjacent cells. This local balancing of flow means that the distribution of concentrations is stable both in space and in time. In other words, if we look at the graph of concentration versus spatial position in figure  \ref{fig:network}(b), of which figure \ref{fig:network}(e) represents a cross section, this profile never changes. This assumption could be biologically realistic for some groups that remain relatively stationary in time, such as animals forming a lek or nesting in colonies. In many animal groups, such as flocks and shoals, however, the spatial distribution of a group is likely to change in time, while maintaining some conserved statistical properties. As an example, the one-dimensional distribution in figure \ref{fig:network}(e) can be summarised by its histogram (fig. \ref{fig:network}(f)) and we can imagine that this histogram is `stable' over time, while the spatial position of zones with different concentration is variable. The stability of the histogram means that the flow of individuals from areas with concentration $C_i$ to areas with slightly larger concentration $C_j$ is the same as the flow in the opposite direction. Given the histogram, there are $N_i$ areas with concentration $C_i$ and $N_j$ areas with concentration $C_j$, which in a similar way as for equation \ref{eq:Weber_balancing_flows} gives the following individual-level probability of moving towards a zone of higher concentration:
\begin{equation}
p(C_i \rightarrow C_j) = \frac{D (N_j C_j - N_i C_i)}{N_i C_i}
\label{eq:Weber_balancing_flows_and_densities}
\end{equation}
which is similar to Weber's law as long as $N_i \simeq N_j$, but would deviate from Weber's law when the histogram changes sharply. It would be impossible to generalise here our approach to more complex definitions of stable groiups, including to flocks and schools, because this would require coming up with relevant definitions of `local density' for a group that is translating in space, and also addressing the issue that many existing models of these systems predict that phase transitions would occur between different flocking modes (see e.g. \cite{Gregoire2003157}). In other words, the diffusion coefficient might be itself density-dependent in these groups.

\section{Long term dynamics of iterated diffusion and anti-diffusion}
\label{sec:longterm}
% MATLAB:script_long_term_dynamics.m
While Weber's law based responses to densities can counteract diffusive forces and are likely to be consistently observed in stable groups with an internal dynamics, can density responses based on Weber's law alone support group stability? 
Because of the linearity of the diffusion and anti-diffusion equations (eq. \ref{eq:diffusion_continuous} and \ref{eq:antidiffusion_continuous}), a combination of diffusion and anti-diffusion has only one stationary state with homogeneous density. The homogeneous density state is stable -and the group disperses across the environment- when diffusion is stronger than anti-diffusion (when $D > \gamma$). When instead the anti-diffusion prevails ($\gamma > D$), the homogeneous state becomes unstable and the model predicts the formation of explosively larger groups. In particular, the dynamics of the aggregation is such that the higher spatial frequencies (small spatial scales / small group sizes) are amplified or attenuated faster than lower frequencies (see Electronic Supplementary Material for an explanation). An important consequence of this is that as soon as the regulation mechanism is noisy, the high frequency noise is quickly amplified.
%\subsection*{Immediate response of individuals}
%\textcolor{red}{First I say what happens with immediate reaction times}
%Let's build the complete model integrating diffusion and anti-diffusion.
%
%\begin{equation}
%\frac{\partial C}{\partial t} = +D \frac{\partial^2 C}{\partial x^2} - \gamma \frac{\partial^2 C}{\partial x^2}
%\label{eq:diff_antidiff_continuous}
%\end{equation}
% This implies an immediate response of individuals to the diffusion event, so that in practice it is equivalent to a slower diffusion (if $D>\gamma$) or slow anti-diffusion (if $\gamma > D$).

\begin{figure*}[bt]
\centering
\includegraphics[width=1\textwidth]{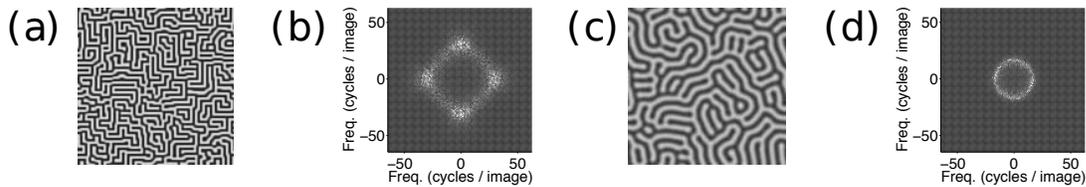}
\caption{
{\bf Pattern formation from iterated diffusion and anti-diffusion}. \textbf{(a)} and \textbf{(c)} Spatial patterns produced on a 128x128 lattice after 15000 iterations of the long-term dynamics model; the starting distribution is random white noise. In both figures $\gamma = 0.9$. In (a) $D=0.13$, while in (b) $D=0.17$. After multiple iterations of the simulation steps, a pattern emerges at a single spatial scale. This is clearly visible in figures \textbf{(b)} and \textbf{(d)}, which plot the amplitude spectrum of the Fourier transform of the patterns in (a) and (c), respectively.
}
\label{fig:sim_start_from_uniform_random}
\end{figure*}

It is important to remember that in making the considerations above we were relying on two assumptions. The first assumption is that diffusion and anti-diffusion take place simultaneously and the gradient climbing response is perfectly accurate. In reality, if the anti-diffusion response is not implemented immediately the anti-diffusion will take place on a distribution that has already been altered by some uncompensated diffusion. Similarly, if the gradient climbing response is noisy, i.e. it also diffuses a bit, this will also result in some uncompensated diffusion.
The second assumption is that local densities can grow with no upper limit. In real world aggregation phenomena, however, densities will reach a saturation at some point. 
Here, we incorporate these additional elements in a discrete model of diffusion and anti-diffusion. The model uncouples in time diffusion and anti-diffusion, and implements a saturating anti-diffusion response to prevent densities from increasing above a saturation point.
We ran computer simulations in which an initial density distribution is altered by alternating steps of diffusion and anti-diffusion in discrete time steps. The first step is a simple diffusion step:
\begin{equation}
\begin{split}
C(x, y, t+1) = C(x, y, t) \\
+ D \Laplace\left(C(x,y,t)\right)
\end{split}
\label{eq:iteratedEq_diffusion_step}
\end{equation}
where
\begin{equation}
\begin{split}
\Laplace \left( C(x,y,t) \right) = - 4 C(x,t) + C(x+1, y, t) + \\
C(x-1, y, t) + C(x, y+1, t) + C(x, y-1, t)
\end{split}
\label{eq:second_step_iterated_antidiffusion}
\end{equation}
The anti-diffusion step is similar to all previous examples, but we also include an additional term ($C(x, y, t+1) \left[1 - C(x, y, t+1)\right]$) to keep densities within the range from zero to one:
\begin{equation}
\begin{split}
C(x, y, t+2) = C(x, y, t+1) \\
- \gamma \Laplace\left(C(x,y,t+1)\right) C(x, y, t+1) \left[1 - C(x, y, t+1)\right]
\end{split}
\label{eq:iteratedEq_antidiffusion_step}
\end{equation}
In the examples shown in figure \ref{fig:sim_start_from_uniform_random} the initial condition is a uniform random distribution and the distribution shown corresponds to 15000 simulation steps.

Under these conditions, we do observe the appearance of spatial patterns of a characteristic scale, at least for some parameters (see fig. \ref{fig:sim_start_from_uniform_random}). While our particular implementation of this long-term model may lack biological realism, for instance in relation to the temporal separation between diffusion and anti-diffusion periods, alternative models that also incorporate a reaction time in the response of individuals are likely to predict the formation of patterns with a characteristic scale: it is implicit in the nature of the diffusion equation that the introduction of events with a characteristic temporal scale -for instance a fixed reaction time in the response to neighbours, also leads to the appearance of spatial patterns with a characteristic spatial scale. % In this case pattern formation happens because anti-diffusion operates on a distribution that has already been altered by diffusion and consequently the scale of the pattern increases with the diffusion coefficient (correspondingly, a bi-laplacian term appears by combining the equations of the model \ref{eq:iteratedEq_diffusion_step} and \ref{eq:iteratedEq_antidiffusion_step}).

\section{Discussion}
% Homeostasis is a central property of biological systems which allows them maintaining -and often recovering- a stable organisation in spite of changes in external conditions and in spite of the internal dynamics of constituent units.

Gregarious animals can form non-homogeneous density distributions and aggregations that persist over time scales typically much longer than the speed of dispersion and movement of individual animals. In our study we show that group stability constrains the individual-level interaction rules that mediate group cohesion, which are required to follow Weber's law.

Weber's law describes a general functioning principle of sensory systems. Traditionally, Weber's law based sensory perception has been discussed in relation to the mechanisms and to the constraints of information processing in the brain \cite{jacob2012relating,van1992theory}. Here, we have shown that Weber's law is also an essential property of collective interactions in stable animal groups.

Sensory perception plays a fundamental role in guiding the interactions of an animal with its environment \cite{akre2014psychophysics}, so it can reasonably be expected that the fundamental properties of sensory systems also play a role in shaping inter-individual interactions and collective behaviour. Previous studies have already established the relevance of decision-making mechanisms based on Weber's law for explaining collective decision-making in ants, bees and fish \cite{Perna2012,von2014pheromone,reina2018psychophysical,Arganda2012}. Theoretical studies also indicate that Weber's law response results from optimal pooling of group-level information for collective decision-making \cite{Arganda2012}.

In the present study, we do not make any assumptions about the mechanisms or about the optimality of the decision-making rules in individual animals. Our only assumption is that animal groups remain stable over long time scales, compared to the time scale of individual animal movements. We further assumed that in the absence of interactions group dynamics would be described by diffusion. Because of the generality of these assumptions, we can conclude that Weber's law response to animal densities is a general property of all stable groups.

We should however be careful not to conclude that all animal aggregation phenomena depend on Weber's law based responses to neighbour densities. Many animal aggregations can be produced also in the complete absence of social interactions and of gregarious behaviour. This is the case for instance when animals are attracted to an environmental feature such as a water source or a particular type of vegetation. In this case, group stability would be predominantly mediated by the external attractor. Yet, our analysis shows that if we -correctly or erroneously- consider that the attraction responses are directed to other group members and not to the environmental feature, whenever the group density is stable we would observe a response to neighbour densities in the form of Weber's law. In some species, the rules of interaction could potentially be extremely complicated, and involve both inter-individual attraction and repulsion. However, as long as the overall effect of these interactions is that of keeping the group stable, we would observe a Weber's law type of response to neighbour densities.

Our model is not a morphogenetic model: the type of response to neighbour densities based on Weber's law that we describe here cannot be used to predict the size or the shape of the groups formed by animals of a particular species. We argue that this is an inevitable consequence of the `universality' of Weber's law: animal groups present a huge variation of size and shape across species which is unlikely to be accounted for by a single shared perceptual rule. In addition, the size of animal groups is typically widely distributed also within one single species, suggesting that its regulation depends on other factors, such as merge and split phenomena, rather than being determined by sensory responses alone \cite{Bonabeau1999a,Niwa2004,ma2011first}. 

While we do actually show that an aggregation model based on Weber's law can lead to the formation of patterns at a fixed scale, this scale is mainly determined by the delay that we introduced between diffusion and anti-diffusion. We could try to find a resemblance between the patterns produced in our long-term iteration model and the spatial distribution of individuals in certain animal groups, such as for instance mussel beds\cite{vandeKoppel2008}. However, we think that it is safer to argue that these pattern formation phenomena leading to a small characteristic spatial scale have little biological relevance in general, for instance because at such small spatial scales the positioning of individuals is more affected by direct individual to individual interactions than by the generic Weber's law response that we consider here.

Paradoxically, while we predict that individual-level interaction responses based on Weber's law should be observed in stable animal groups, Weber's law alone cannot explain the formation of groups of a particular form or size and is not sufficient by itself to explain the stability of pre-existing groups (because of reaction times and amplification of noise).We can conclude that Weber's law based interactions are a necessary but not entirely sufficient feature of stable group.

Our study points to a general relation between Weber's law and collective behaviour. While Weber's law can explain several perceptual phenomena, there are many instances in which sensory perception deviates from Weber's law behaviour.  Future studies should try to relate also deviations from Weber's law, as well as other perceptual phenomena, to collective behaviour. For example, many sensory stimuli remain undetected when their intensity falls below a perception threshold. In the context of social interactions, perceptual thresholds for responding to conspecifics can directly affect the way animals rely on private vs. social information for taking decisions.

Our approach for deriving interaction responses in animal groups is complementary to other studies based on direct observations of interacting individuals. While direct observations of behaviour inform us about how animals of a particular species interact, our approach informs us about which individual-level interactions are `unavoidable' given a particular collective phenomenon. In this particular case, individual-level responses in the form of Weber's law are unavoidable in stable density groups. This `inverse problem' approach has been previously applied to the study of other collective behaviour phenomena (e.g. \cite{perna_duality}, \cite{Mann_entropy}) and we are confident that in future our approach can be extended to the study of also other more complex forms of group coordination.

\section{Author Contributions}
Formulated the initial research idea and approach: AP; discussed various modelling approaches: AP; JLD; developed the non-linear Weber's law models: GF; implemented the computer based simulations: AP; wrote the article: AP. All authors were involved in discussions on different aspects of the study.

\section{Acknowledgements}
JLD is a Research Director from the Belgian National Fund for Scientific Research (FRS-FNRS).

\section{Data Accessibility}
This article has no data. The code used in the simulations is available at \\https://github.com/pernafrost/weber\_law\_antidiffusion

\section{Funding Statement}
AP was supported during part of this study by a fellowship from the Belgian National Fund for Scientific Research (FRS-FNRS)\\
GF is supported by a Newton International Fellowship of the Royal Society.

% \bibliography{/Users/perna/articoli/bibliografia.bib}

\end{document}